\documentclass[11pt,tightenlines,aps,onecolumn,prd,showpacs,nofootinbib,showkeys,superscriptaddress,preprintnumbers]{revtex4-1}

\usepackage{amsmath}
\usepackage{graphicx}
\usepackage{color}
\newcommand{\apjl}{ApJL}

\newcommand{\mnras}{MNRAS}

\newcommand{\araa}{Ann. Rev. Astron. Astropys.}
\newcommand{\pasa}{Pub. Astron. Soc. Aust.}

\newcommand{\aap}{Astron. Astrophys.}

\newcommand{\baas}{Bull. Am. Astron. Soc}

\newcommand{\um}{$\mu$m}
\newcommand{\mic}{$\mu$m}

\makeatother

\begin{document}
\title{Cosmological advection flows in the presence of primordial black holes as dark matter and formation of first sources}

\author{A. Kashlinsky}
\affiliation{Code 665, Observational Cosmology Lab, NASA Goddard Space Flight Center, 
Greenbelt, MD 20771 and
SSAI, Lanham, MD 20770; email: Alexander.Kashlinsky@nasa.gov} 


\def\plotone#1{\centering \leavevmode
\epsfxsize=\columnwidth \epsfbox{#1}}

\def\wisk#1{\ifmmode{#1}\else{$#1$}\fi}

\def\wm2sr {Wm$^{-2}$sr$^{-1}$ }		
\def\nw2m4sr2 {nW$^2$m$^{-4}$sr$^{-2}$\ }		
\def\nwm2sr {nWm$^{-2}$sr$^{-1}$\ }		
\def\nw2m4sr {nW$^2$m$^{-4}$sr$^{-1}$\ }
\def\Ncut {$N_{\rm cut}$\ }
\def\lt     {\wisk{<}}
\def\gt     {\wisk{>}}
\def\le     {\wisk{_<\atop^=}}
\def\ge     {\wisk{_>\atop^=}}
\def\lsim   {\wisk{_<\atop^{\sim}}}
\def\gsim   {\wisk{_>\atop^{\sim}}}
\def\kms    {\wisk{{\rm ~km~s^{-1}}}}
\def\Lsun   {\wisk{{\rm L_\odot}}}
\def\Msun   {\wisk{{\rm M_\odot}}}
\def\um     { $\mu$m\ }
\def\sig    {\wisk{\sigma}}
\def\etal   {{\sl et~al.\ }}
\def\eg	    {{\it e.g.\ }}
\def\ie     {{\it i.e.\ }}
\def\bsl    {\wisk{\backslash}}
\def\by     {\wisk{\times}}
\def\cosec {\wisk{\rm cosec}}
\def\mic {\wisk{ \mu{\rm m }}}

\def\amin   {\wisk{^\prime\ }}
\def\asec   {\wisk{^{\prime\prime}\ }}
\def\cc     {\wisk{{\rm cm^{-3}\ }}}
\def\deg     {\wisk{^\circ}}
\def\ddeg   {\wisk{{\rlap.}^\circ}}
\def\damin  {\wisk{{\rlap.}^\prime}}
\def\dasec  {\wisk{{\rlap.}^{\prime\prime}}}
\def\approxeq{$\sim \over =$}
\def\abouteq{$\sim \over -$}
\def\percm{cm$^{-1}$}
\def\percmsq{cm$^{-2}$}
\def\percmcub{cm$^{-3}$}
\def\perhz{Hz$^{-1}$}
\def\perpc{$\rm pc^{-1}$}
\def\persec{s$^{-1}$}
\def\peryr{yr$^{-1}$}
\def\te{$\rm T_e$}
\def\tenup#1{10$^{#1}$}
\def\to{\wisk{\rightarrow}}
\def\thin{\thinspace}
\def\uk{$\rm \mu K$}
\def\p{\vskip 13pt}

\begin{abstract}
In the inflation-based  cosmology the dark matter (DM) density component starts moving with respect to the universal expansion at $z_{\rm eq}\sim 3,200$ while baryons remain frozen until $z_{\rm rec}\sim 1,100$. It has been suggested that in this case post-linear corrections to the evolution of small fluctuations would result, for the standard $\Lambda$-dominated cold DM (CDM) model, in delayed formation of early objects as supersonic advection flows \textcolor{black}{develop after recombination}, so baryons are not immediately captured by the DM gravity o\textcolor{black}{n} small scales. We develop the hydrodynamical description of such two-component advection and show that, in the supersonic regime, the advection within irrotational fluids is governed by the gradient of the difference of the kinetic energies of the two (DM and baryonic here) components. We then apply this formalism to the case where DM is made up of LIGO-type black holes (BHs) and show that there the advection process on scales relevant for early structure collapse will differ significantly from the earlier discussed (CDM) case because of the additional granulation component to the density field produced during inflation. The advection here will lead efficiently to the common motion of the DM and baryon components on scales relevant for collapse and formation of first luminous sources. \textcolor{black}{This leads to early collapse making easier to explain the existence of supermassive BHs observed in quasars at high $z>7$. The resultant net advection rate reaches minimum around $\lsim 10^9M_\odot$ and subsequently rises to a secondary maximum near the typical mass of $\sim 10^{12}M_\odot$, which may be an important consideration for formation of galaxies at $z\lsim$(a few)}.
\end{abstract}

\maketitle

\section{Introduction}

The linear Newtonian growth theory of cosmic structures is now established \citep[e.g.][]{Harrison:1967}. 
However, 
\cite{Tseliakhovich:2010} noted the importance of advection flows in post-linear approximation for standard cosmological model, where the dark matter (DM) density field  approaches the power $P\propto k^{-3}$ at large $k$, while preserving the initial Harrison-Zeldovich shape $P\propto k$ on scales, $\sim k^{-1}$, exceeding the horizon at matter-radiation equality, redshift $z_{\rm eq}\simeq 3200$, when the DM component starts growing. This results in a coherent velocity field on scales $\lsim$ a few Mpc corresponding to early collapsing structures. At the same time baryons remain frozen into the comoving (radiation) frame until recombination, $z_{\rm rec}\simeq 1100$, when DM is already moving supersonically \textcolor{black}{relative to baryon's sound speed} \citep[][]{Ricotti:2008}. Baryons then find themselves moving supersonically in {\it highly coherent} DM flows  likely delaying their collapse to form luminous objects \cite{Tseliakhovich:2010}. Much work followed on this potentially important effect \citep[e.g.][]{Maio:2011,Stacy:2011,Greif:2011,Ahn:2016,
Ahn:2018,Blazek:2016,Schmidt:2016,Hirata:2018}.

The advection hydrodynamics would be modified if LIGO-mass primordial black holes (PBHs) make up DM. The possibility was proposed to explain their apparent merger rate \citep{Bird:2016,Clesse:2017a}, or the source-subtracted cosmic infrared background (CIB) \citep{Kashlinsky:2016}, where this conjecture naturally reproduces the amplitude and shape of the earlier uncovered near-IR  source-subtracted CIB fluctuations \citep{Kashlinsky:2005a} and their strong spatial coherence with cosmic X-ray background implying populations containing substantial BH proportions \citep{Cappelluti:2013}. See review \cite{Kashlinsky:2018}. Following the first LIGO detection during its short engineering test run \citep{Abbott:2016c} the O1+O2 run uncovered 10 significant BH mergers of $\sim10-50M_\odot$ masses with low-to-zero spins \citep{The-LIGO-Scientific-Collaboration:2019}. 
The O3/O4 LIGO runs at increasing sensitivity should provide critical insights into the possible PBH-DM collusion. Theoretical mechanisms for such PBHs are discussed in  \cite{Garcia-Bellido:1996,Jedamzik:1997,Musco:2005,Carr:2016,Sasaki:2018}. This decade's new EM-based efforts should shed critical light on the PBH-DM linkage \citep{Kashlinsky:2019b}, particularly in the CIB realm \citep{Kashlinsky:2019,Kashlinsky:2019a}.  Recent discussions of cosmogonical implications of this conjecture include \citep{Carr:2018,Inman:2019,Hasinger:2020}.

If PBHs constitute DM their granulation produces an additional power component \citep{Meszaros:1974,Meszaros:1975} of shot-noise type on scales beyond the horizon scale at the time of their formation \citep{Kashlinsky:2016}, which modifies the advection mechanism compared to  \cite{Tseliakhovich:2010}. Here we show that the advection in the presence of PBHs as DM would be efficient in early \textcolor{black}{equalizing} the velocity fields of DM and baryons. This enables a sufficiently early formation of the BH seeds to explain the existence of supermassive BHs (SMBHs) observed in quasars at $z\gsim 7$. The advection rate resulting from the $\Lambda$DM and PBH components is such that the \textcolor{black}{equalizing} the two velocity fields becomes less efficient for total halo masses in the $10^9-10^{11}M_\odot$ range and peaks up again for haloes around a few times $10^{12}M_\odot$, the typical masses of modern galaxies potentially affecting/determining formation of structures on galactic mass-scales.

We assume two  primordial density field components: 1) from the standard inflationary era ($\Lambda$DM), and 2) the PBH granulation fluctuation component generated after   inflation. We adopt DM fraction, $f_{\rm PBH}\leq 1$, for PBHs; the results are scaleable with $f_{\rm PBH}$. 
\citep{Lacki:2010} argued  that PBHs must either make all the DM or contribute (almost) nothing since otherwise they accrete particle DM producing highly luminous annihilation sources contradicting $\gamma$-ray observations. 
\section{Cosmological advection flows}

\textcolor{black}{We use hydrodynamical descriptions for the common evolution of the PBH-DM and baryon components. While this is obviously applicable to baryons, some caveats are in order for the PBHs. If PBHs of mass $M_{\rm PBH}$ contribute the fraction $f_{\rm PBH}$ to the the average density, their mean comoving separation is $\bar{r}_{\rm PBH}\simeq 0.44 f_{\rm PBH}^{1/3}(M_{\rm PBH}/30M_\odot)^{-1/3}$Kpc and the number of PBHs contained in comoving scale $r$ is $N_{\rm PBH}\simeq(r/\bar{r}_{\rm PBH})^3\gg1$ on scales of relevance here. Provided we consider scales encompassing $N_{\rm PBH}\gg1$ their evolution is described by the stellar-dynamical Jeans equations, which, absent stellar-dynamical pressure (in virialized stellar systems), are equivalent to Euler's fluid equations; we call both the ``Euler-Jeans" equations as they are derived from moments of Boltzman's equation.} \textcolor{black}{We consider comoving scales $r\gsim0.03$Mpc, so the Euler-Jeans equations are valid provided $M_{\rm PBH}\ll 10^7 f_{\rm PBH}(r/0.03{\rm Mpc})^3M_\odot$; this range covers the LIGO-type PBH masses examined throughout.} \textcolor{black}{The PBH mass-range considered in the analysis here is comfortably within the $1\!-\!\sigma$ upper limits on the granulation power, discussed below, from the Lyman-forest observations and simulations \cite{Afshordi:2003,Murgia:2019}.}

We follow standard notations for the Newtonian evolution of density fluctuations and their flows 
for the two components, ``d" (DM) and ``b" (baryons). After recombination, the Euler-Jeans equations for their evolution are given by eqs 6 of \cite{Tseliakhovich:2010}:
\begin{equation}
 \dot{\boldsymbol{v}}_{\rm d,b} +H\boldsymbol{v}_{\rm d,b} + a^{-1}
(\boldsymbol{v}_{\rm d,b} \cdot\boldsymbol{\nabla})\boldsymbol{v}_{\rm d,b} =-a^{-1}\boldsymbol{\nabla}\phi -\frac{a^{-1}}{{\rho}_{\rm d,b}}\boldsymbol{\nabla}p_{\rm d,b}
\label{eq:euler}
\end{equation}
where $a=(1+z)^{-1}, H(z)=\dot{a}/a$, $\phi$ is the gravitational potential and $p$ is the presure in each component.  We use the Lamb transformation $(\boldsymbol{v}\cdot\boldsymbol{\nabla})\boldsymbol{v}=\frac{1}{2}\boldsymbol{\nabla}(v^2)- \boldsymbol{v}\times\boldsymbol{\omega}$ where vorticity $\boldsymbol{\omega}=\boldsymbol{\nabla}\times\boldsymbol{v}$ and $v=|\boldsymbol{v}|$ \citep{Lamb:1975}. Gravity is a potential force  inducing irrotational flows so vorticity is small in linear regime \citep{Peebles:1969,Efstathiou:1979}
; thus $(\boldsymbol{v}\cdot\boldsymbol{\nabla})\boldsymbol{v}=\frac{1}{2}\boldsymbol{\nabla}(v^2)$. 

The relative  baryons--DM velocity is important for collapsed structure formation. Once the two components move together, the gas can collapse in the formed DM halos and fragment subject to its cooling and fragmentation efficiency \citep{Hoyle:1953,Rees:1977,Kashlinsky:1982}. Since both flows are driven by the same gravitational potential $\phi$, subtracting ``d" from ``b" components in (1) gives, for irrotational flows:
%
\begin{equation}
\frac{\partial }{\partial t}(a{\boldsymbol{V}}_{\rm bd})= -\boldsymbol{\nabla}({\cal K}_{\rm b}-{\cal K}_{\rm d} + c_s^2\delta_b) 
\label{eq:advection_final}
\end{equation}
where $\boldsymbol{V}_{\rm bd}\equiv \boldsymbol{v}_{\rm b}-\boldsymbol{v}_{\rm d}$, each component's kinetic energy per unit mass ${\cal K}=\frac{1}{2}v^2$,  $c_s^2\equiv\frac{\partial p_{\rm b}}{\partial\rho_{\rm b}}$ is the adiabatic sound speed squared, assumed uniform, and DM is taken to be pressurerless.  Eq. \ref{eq:advection_final} with the continuity equation for DM later are equivalent to equations (6) of \cite{Tseliakhovich:2010} when baryons comove with DM. 
 DM dominates the peculiar gravity and starts moving at matter-radiation equality, $z_{\rm eq}\simeq 3,200$, whereas baryons 
 \textcolor{black}{start growing} with $\boldsymbol{v}_b\simeq 0$ at $z_{\rm rec}\simeq 1,090$. Both components move supersonically after recombination \cite{Ricotti:2008,Tseliakhovich:2010}. 
Relevant solutions/consequences of (\ref{eq:advection_final}) are:

\begin{enumerate}
\item Eq. \ref{eq:advection_final} provides  an {\it exact} description of the relative motions of the irrotational flow components \textcolor{black}{until the DM shell-crossing}. For supersonic flows under the same potential force (gravity) the advection is influenced by the gradient of the difference in kinetic energies, ${\cal K}$, of the flow components, driving their (kinematic) mixing.
\item In steady-state $\boldsymbol{v}_{\rm b}=\boldsymbol{v}_{\rm d}$, i.e. baryons and DM move coherently on all scales (to within $c_s$).  This solution always exists even at time/space-varying $\boldsymbol{v}$ in the linear approximation, when the RHS of (\ref{eq:advection_final}) vanishes. This regime is reached as the result of advection.
\item The general solution to eq.\ref{eq:advection_final} is the sum of two parts, the first of which is $\boldsymbol{V}_{\rm bd}(z) = -\boldsymbol{v}_{\rm d}(z\!\!=\!\!1000)(1+z)/1000$ if $v_{\rm b}=0$ initially. This solution fully describes  the component of $\boldsymbol{V}_{\rm bd}$ perpendicular to $\boldsymbol{\nabla}({\cal K}_{\rm d})$. When it dominates advection flows may suppress/delay the onset of baryonic collapse \cite{Tseliakhovich:2010}.
\item Because $v_{\rm d}$ grows with time, the advection term on the RHS of eq. \ref{eq:advection_final} may become important and also more efficient in decreasing $\boldsymbol{V}_{\rm bd}$. 
The remaining relative baryon-DM velocity, $\boldsymbol{V}_{\rm bd}$, is then dominated by the solution along the \textcolor{black}{direction} defined by $\boldsymbol{\nabla}({\cal K}_{\rm d})$. If $r$ denotes the coordinate along this \textcolor{black}{direction}, the baryons eventually catch up with the motion of DM, i.e. $\dot{V}_{\rm bd}>0$, if $\partial (v_{\rm d}^2)/\partial r<0$ ($v_{\rm d}^2$ decreases with increasing distance) and vice versa. The rate at which baryons catch up with the moving DM is given by the gradient of the kinetic energy of the DM bulk flow, $\partial {\cal K}_{\rm d}/\partial r$.
\item Evolution of $\boldsymbol{V}_{\rm bd}$, (\ref{eq:advection_final}), along the line defined by $\boldsymbol{\nabla}({\cal K}_{\rm d})$ starts at $\boldsymbol{V}_{\rm bd}=-\boldsymbol{v}_{\rm d}$ proceeding with baryons catching up with DM ($\boldsymbol{v}_{\rm b}=\boldsymbol{v}_{\rm d}$) on advection timescale $t_{\cal A}\sim L_{\cal A}/v_{\rm d}$ where $L_{\cal A}=a\textcolor{black}{v_{\rm d}^2}/|\boldsymbol{\nabla}({\cal K}_{\rm b}-{\cal K}_{\rm d})|$; the rate at which velocities are equalized, $v_{\rm d}/t_{\cal A}$, being closely related to the advection rate defined below. Consequently, for the  $\Lambda$DM power spectrum the advection terms can become comparable to or larger than the expansion time \cite[Sec. II.C in][]{Tseliakhovich:2010}. If the power spectrum (e.g. PBH-DM) is such that $t_{\cal A}$ wins over the expansion \textcolor{black}{time} the balancing of the b,d flows proceeds more efficiently, similar to analogous problems in atmospheric advection \cite{Vasilkov:1975}.
\item When $v_{\rm b}=0$ to within $c_s$ to begin with, but $v_{\rm d}\gg c_s$ so that $v_{\rm b}^2\ll v_{\rm d}^2$, we can Fourier transform velocities, 
 rewriting (\ref{eq:advection_final}) 
\textcolor{black}{
\begin{equation}
\frac{\partial[a\boldsymbol{U}_{\rm bd}(\boldsymbol{k})]}{\partial t}\propto  \boldsymbol{k} \int U_{\rm d}(\boldsymbol{k}-\boldsymbol{k}_1)U_{\rm d}(\boldsymbol{k}_1)d^3\boldsymbol{k}_1
\label{eq:mode-coupling}
\end{equation}
}
where $\boldsymbol{U}_{\rm bd}$ is the Fourier mode of $\boldsymbol{V}_{\rm bd}$, and the RHS integral represents the convolution of $U_d\equiv|\boldsymbol{U}_{{\rm d}, \boldsymbol k}|$ with itself, $U_{{\rm d}, \boldsymbol k} \star U_{{\rm d}, \boldsymbol k}$.
Consequently, Fourier harmonics no longer evolve independently \cite[e.g.][]{Efstathiou:1990}.
\item  In the presence of rotation/vorticity the RHS of eq. \ref{eq:advection_final} contains the additional term $(\boldsymbol{v}_{\rm b}\times\boldsymbol{\omega}_{\rm b} - \boldsymbol{v}_{\rm d}\times\boldsymbol{\omega}_{\rm d})$. 
\textcolor{black}{\item When $c_s\ll v_d$ and $v_{\rm b}=0$ at recombination the relative baryon-DM velocity evolves as
\begin{equation}
\boldsymbol{V}_{\rm bd}=  -\frac{1+z}{1+z_{\rm rec}} \boldsymbol{v}_{\rm d,rec} -a^{-1} \int_{t(z_{\rm rec})}^{t(z)} \boldsymbol{{\cal A}}_{\cal K}\; a dt
\label{eq:v_final}
\end{equation}
where $\boldsymbol{{\cal A}}_{\cal K} \equiv - a^{-1}\boldsymbol{\nabla}{\cal K}_{\rm d}$ is the local advection rate. When the latter dominates the first term on the RHS of eq. \ref{eq:v_final} the advection speeds up the equalizing of the baryon-DM velocity. Eq. \ref{eq:v_final} is correct to $O(v_{\rm b}^2/v_{\rm d}^2)$ leading to steady-state at $V_{\rm bd}=0$.
}
\end{enumerate}

Below we adopt: $h\equiv H_0/$(100 km/sec/Mpc)$=0.7,\Omega_{\rm d}h^2=0.11, \Omega_{\rm b}h^2=0.023, \Omega_{{\rm m},0}=\Omega_{\rm d}+\Omega_{\rm b}=0.3,\Omega_\Lambda+\Omega_{\rm d}+\Omega_{\rm b}=1, \sigma_8=0.9$. At $z_{\rm rec} \gg z\gg1$, $\Omega_{\rm m}(z)=\Omega_{{\rm m},0}(1+z)^3/[\Omega_{{\rm m}, 0}(1+z)^3+\Omega_\Lambda]\simeq 1$, the Hubble constant $H(z)\simeq H_0 \sqrt{\Omega_{\rm m,0}}(1+z)^{3/2}$, $t_{\rm cosm}(z)\simeq\frac{2}{3} H^{-1}(z)\simeq 0.2(\frac{1+z}{20})^{-3/2}$ Gyr. 

\section{Advection flows for PBH dark matter}


Following \cite{Kashlinsky:2016} we assume that the power spectrum, $P_{\rm m}(k)$, of matter fluctuations responsible for structure formation at recombination, $z\simeq10^3$, is made up of 1) the $\Lambda$DM component from the inflationary period and 2) the component from LIGO-type PBHs contributing a fraction $f_{\rm PBH}$ of the DM:
\begin{equation}
P_{\rm m}(k) = P_{\Lambda{\rm DM}}(k)+P_{\rm PBH}=P_{\Lambda{\rm DM}}(k)+1.2\times10^{-8}f_{\rm PBH}\left(\frac{M_{\rm PBH}}{30M_\odot}\right)\left(\frac{1+z}{1000}\right)^{-2}{\rm Mpc}^3
\label{eq:power}
\end{equation}
The PBH component $P_{\rm PBH}\propto f_{\rm PBH}M_{\rm PBH}$. For an extended PBH mass function \citep[e.g.][]{Clesse:2018,Carr:2018,Inman:2019}  $M_{\rm PBH}$ represents the effective PBH mass\textcolor{black}{, it being the eigen-value after suitably averaging over the PBH mass-function}. 

The advection eq. \ref{eq:advection_final}, when starting at $z_{\rm rec}$ at $\boldsymbol{v}_{\rm b}=0$, must be complemented with the
continuity equations for each component, $\left[\dot{\delta}_{\rm d,b}+a^{-1}\boldsymbol{\nabla}\cdot\boldsymbol{v}_{\rm d,b}\right]=-a^{-1}\boldsymbol{\nabla}\cdot(\delta_{\rm d,b}\boldsymbol{v}_{\rm d,b})$. 
Its general solution is the sum of two terms: $\boldsymbol{v}_d\equiv\boldsymbol{v}_{{\rm d, 1}}+\Delta \boldsymbol{v}_{\rm d}$. The first term satisfies, in this gauge, the sum in square brackets  being $\left[\ldots\right]=0$. Hence $\Delta \boldsymbol{v}_{\rm d}=-\frac{\delta_{\rm d}}{1+\delta_{\rm d}}\boldsymbol{v}_{{\rm d,1}}\simeq-\delta_{\rm d}\boldsymbol{v}_{{\rm d,1}}+O(\delta_d^2 v_{{\rm d,1}})$. Fourier-transforming $\delta_{\rm d}(\boldsymbol{r}), \boldsymbol{v}_{\rm d}(\boldsymbol{r})$ into $\Delta_{\rm d}(\boldsymbol{k}), \boldsymbol{U}_{\rm d}(\boldsymbol{k})$
for the irrotational flow, gives $\boldsymbol{U}_{\rm d}=\boldsymbol{U}_{\rm d,1}+(\boldsymbol{U}_{\rm d,1}\star\Delta_{\rm d})_{\boldsymbol{k}}$, 
where ${\boldsymbol{U}}_{{\rm d,1}} =-i\boldsymbol{k}a\dot{\Delta}_{\rm d}/k^2$  and $\boldsymbol{U}_{{\rm d,1}}\star\Delta_d\equiv\int \boldsymbol{U}_{{\rm d,1}}(\boldsymbol{k}_1)\Delta_{\rm d}(\boldsymbol{k}-\boldsymbol{k}_1)d^3\boldsymbol{k}_1$. 
The small-to-large scale mode-coupling is less pronounced here because of the absence of the $k^{-1}$ weight for $\Delta_{\rm d}$ compared to \textcolor{black}{(\ref{eq:mode-coupling})}. For the white-noise PBH component dominating small-scale power, $\Delta(k)$=const, the convolution integral $\propto \int \boldsymbol{U}_{{\rm d,1}}(\boldsymbol{k}_1)d^3\boldsymbol{k}_1=\boldsymbol{v}_{\rm d,1}(0)$ is constant with $k$, while $U_{\rm d,1}$ increases toward small scales.

\begin{figure}[h!]
\includegraphics[width=7in]{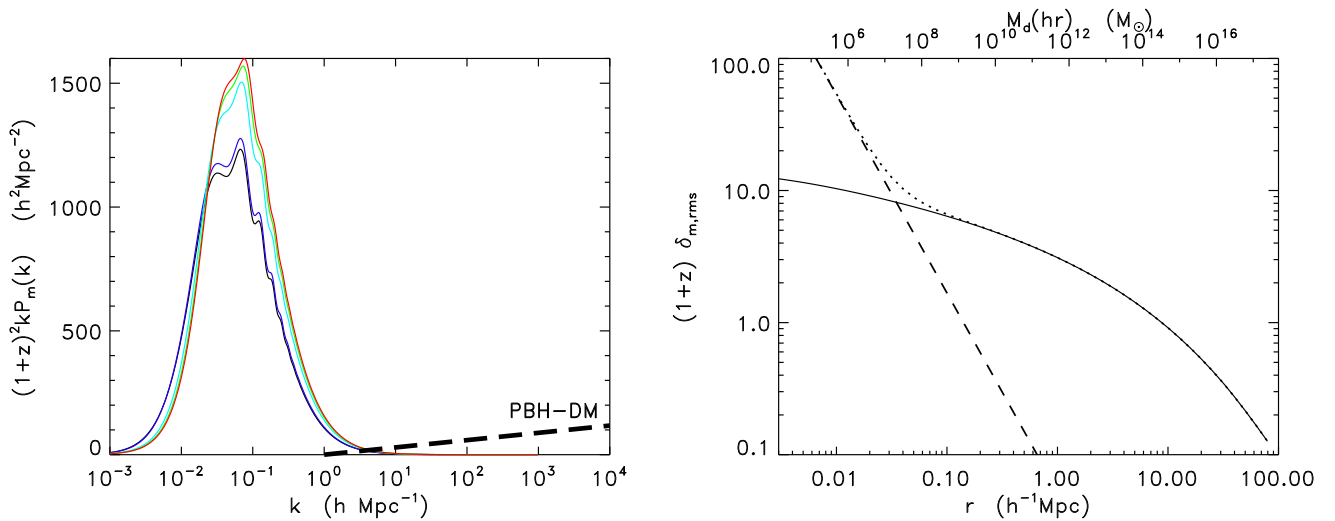}
\caption{{\bf Left}: Logarithmic contribution to the DM velocity variance, $kP_{\rm m}(k)$, times the power growth factor for Einstein-deSitter regime. Lines, for the $\Lambda$DM component, are for $z=1,000$ (red), 900 (green), 700 (indigo), 100 (blue) and 20 (black). Thick dashes show DM-PBH contribution for $f_{\rm PBH}M_{\rm PBH}=30M_\odot$ {\bf Right}: RMS density fluctuation dispersion times the Einstein-deSitter growth-factor 
vs the comoving radius subtending the matter mass in the upper horizontal axis. Solid, dashed, dotted lines show the $\Lambda$DM, DM-PBH term, and their sum.}
\label{fig:fig1}
\end{figure}

Hence, like \cite{Tseliakhovich:2010}, we adopt the linear growth velocity evolution, i.e. $k^{2}P_v(k,z)=\left[\frac{a\dot{\Delta}(k,z)}{\Delta(k,z)}\right]^2P_m(k,z)$. In linear approximation $\Delta(k)\propto a$, but to describe the full non-linear evolution of density fields requires solving (\ref{eq:euler}) \cite{Tseliakhovich:2010} or using the "stable-clustering" approach \citep{Peacock:1996}.  Such corrections are scale-dependent but small (\cite{Tseliakhovich:2010} show in Fig. 2 the $\lsim$O(10\%) non-linear corrections to $P_{\Lambda{\rm DM}}$ at $z\gsim 40$ peaking around $k\simeq100-300$Mpc$^{-1}$) so we take $P_v = H^2(z) [\Omega_{\rm m}(z)]^{1.2}P_{\rm m}/k^2$ with the caveats discussed later. Fig. \ref{fig:fig1},left plots the $\Lambda$DM and PBH  parts of $kP_{\rm m}(k)$ at $z=1000, 900, 700, 100, 20$ for standard cosmological parameters using CAMB\footnote{\url{https://lambda.gsfc.nasa.gov/toolbox/tb_camb_form.cfm}}. Fig. \ref{fig:fig1},right shows the rms density fluctuation, $\delta_{\rm m,rms}^2=\frac{1}{2\pi^2}\int P_{\rm m}W^2(kr)k^2 dk$ from these components, where $W(y)=3j_1(y)/y$. 

\begin{figure}[h!]
\includegraphics[width=7in]{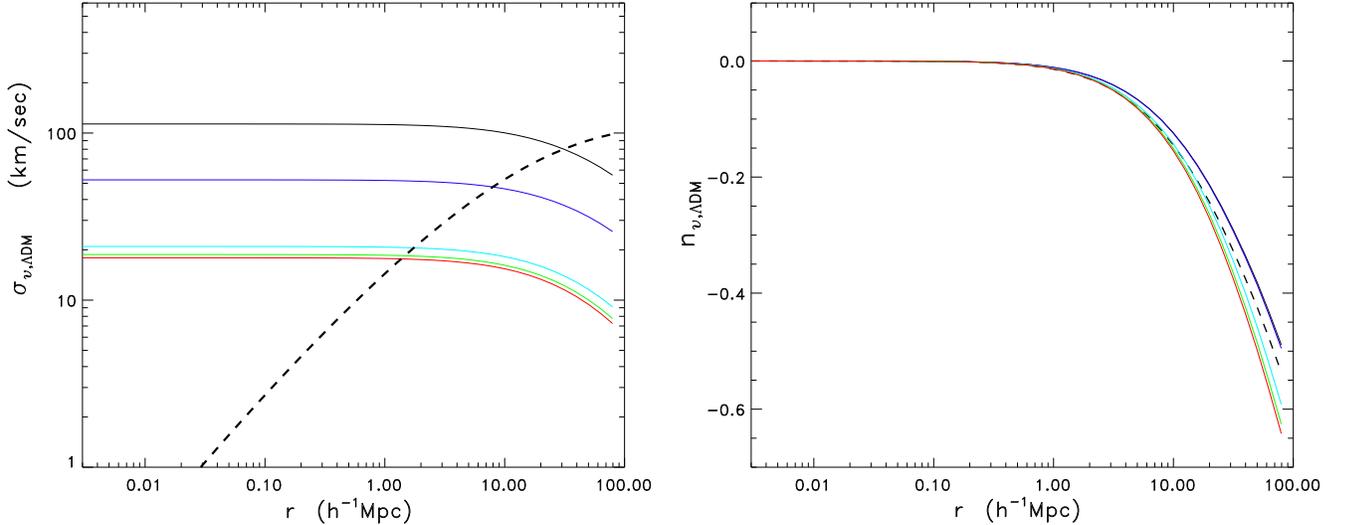}
\caption{Velocity parameters for the $\Lambda$DM component. {\bf Left}: Solid lines  show the velocity dispersion vs comoving scale at different $z$ in the color notation of Fig. \ref{fig:fig1},left. Thick dashed line shows the 1-D relative velocity $v_{\rm rel}(r)\equiv \sqrt{\sigma_v^2(0)-\sigma_v^2(r)}$ at $z=20$. {\bf Right}: Logarithmic slope of the $\Lambda$DM component's velocity dispersion, $n_v$ vs $r$, for the color lines on the left. For PBH component $n_v=-0.5$.}
\label{fig:fig2}
\end{figure}

The mean kinetic energy of the DM component is:
\begin{equation}
\bar{\cal K}_{\rm d} = \frac{1}{2} \sigma_v^2 =\frac{1}{4\pi^2} \Omega_m^{1.2}(z) a^2H^2 \int_0^\infty P_{\rm m}(k)W^2(kr)dk \equiv \frac{1}{2}(\sigma_{v,\Lambda {\rm DM}}^2 + \sigma_{v,{\rm PBH}}^2)
\label{eq:sigma_v}
\end{equation}
where the two RHS terms arise from the two power components, (\ref{eq:power}). The 1-D velocity variance, $\sigma_v^2$, is related to the "dot" velocity correlation function, $\xi_v(\boldsymbol{r})=\langle \boldsymbol{v}(\boldsymbol{r}^\prime)\cdot \boldsymbol{v}(\boldsymbol{r}^\prime+\boldsymbol{r})\rangle$\citep{Peebles:1980,Vittorio:1986,Kashlinsky:1992}; the relative 1-D velocity relevant for turn-around/collapse is $v_{\rm rel}(r)=\sqrt{\sigma_v^2(0)-\sigma_v^2(r)}$. 
Fig. \ref{fig:fig2},left shows the $\Lambda$DM component of the velocity variance, $\sigma_{v,\Lambda{\rm DM}}$. The resultant flow of the $\Lambda$DM component is highly coherent out to comoving $r\sim$ (a few)Mpc. We define the velocity dispersion slope, $n_v\equiv\frac{\partial \ln \sigma_v(r)}{\partial \ln r}=\frac{1}{2}\frac{\partial \ln \bar{\cal K}_{\rm d}(r)}{\partial \ln r}$. On scales where  $\sigma_v$ is highly coherent the relative 1-D velocity is $v_{\rm rel}\simeq \sqrt{-2n_v} \sigma_v(0)$. Fig. \ref{fig:fig2},right plots $n_v$ vs $r$ for the $\Lambda$DM component.  For the PBH component $n_v=-0.5$.
Eq. 2 of \cite{Tseliakhovich:2010} gives $c_s\simeq 5.5-1.4$ km sec$^{-1}$ at $z=1000-100$ with $c_s<0.5$ km/sec at $z<30$ (until reionization/reheating). The acoustic pressure term is subdominant compared to the DM kinetic energy in (\ref{eq:advection_final}) until first sources form and reheat the baryonic gas.

The mean of $\boldsymbol{V}_{\rm bd}\cdot$(eq. \ref{eq:advection_final}) is zero over a finite volume because of the conservation of relative energy. At each point the two components get mixed by the instantaneous gradient of the (difference in) kinetic energy with non-zero rms rate. The rms  measure of the advection rate due to the PBH component with the mean bulk kinetic energy $\bar{\cal K}$ is (with $\int_0^\infty W^2(y)dy\simeq 2$):
\begin{equation}
\bar{{\cal A}}_{\cal K}\equiv - a^{-1}\frac{\partial \bar{\cal K}_{\rm d}}{\partial r} =  -a^{-1} n_{v,\Lambda{\rm DM}} \frac{\sigma_{v,\Lambda{\rm DM}}^2}{r}+ \frac{\Omega_m^{1.2}(z)}{\pi^2} aH^2 P_{\rm PBH}(z) \frac{1}{r^2}
\label{eq:advection_pbh}
\end{equation}
We express the advection rate in km/sec/Gyr, so ${\cal A_K}=1$ \textcolor{black}{equalizes} relative motions of 1 km/sec in 1 Gyr. 

\begin{figure}[h!]
\includegraphics[width=6.25in]{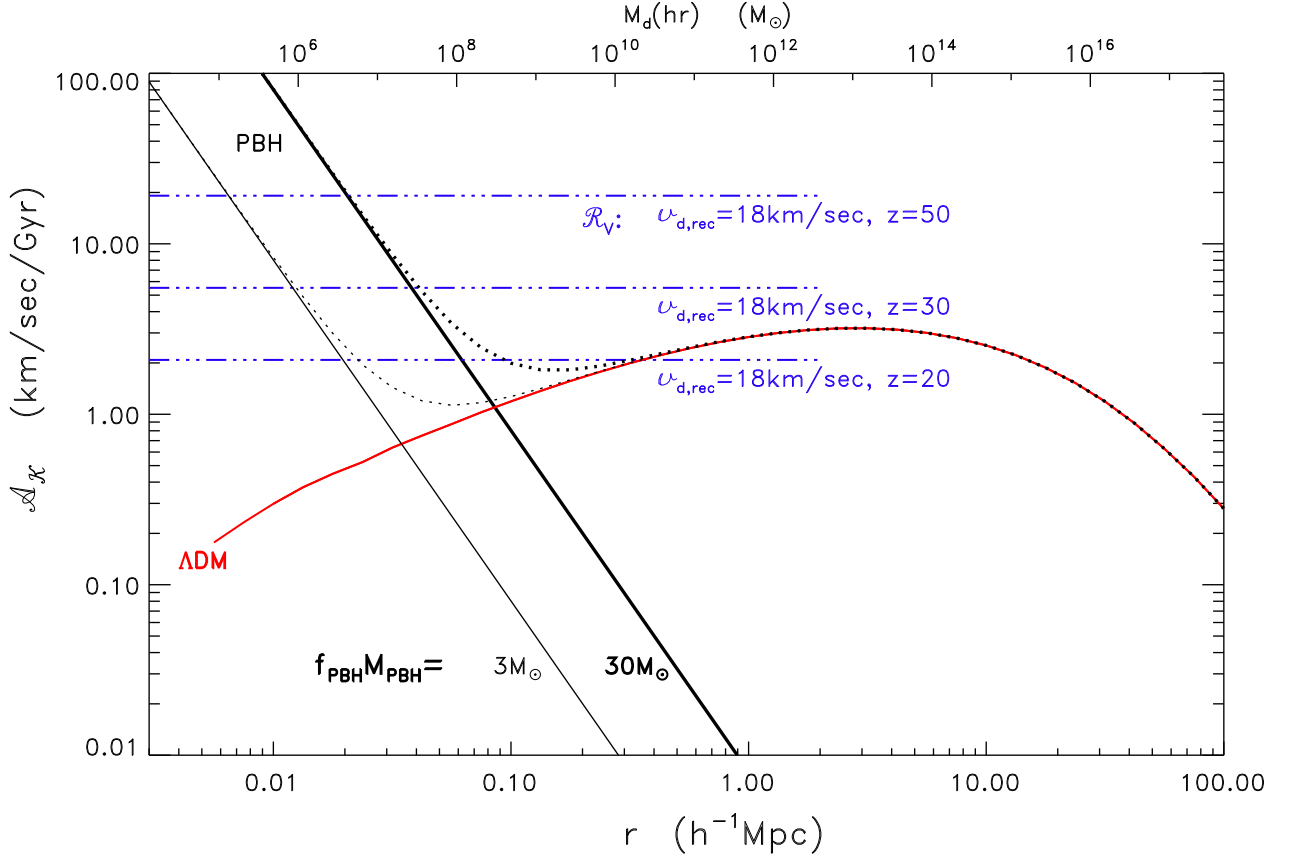}
\caption{The advection rate (solid) 
for the $\Lambda$DM (red) and PBH (black) components with $f_{\rm PBH}M_{\rm PBH}$ marked. Dotted lines show  total ${\cal A_K}$. The advection from the PBH-DM component is more efficient in \textcolor{black}{equalizng} DM and baryonic velocity components than the $\Lambda$DM alone at scales where $\Lambda$DM component starts dominating the density fluctuation per Fig. \ref{fig:fig1},right. The PBH advection rate  continues to be higher than of the $\Lambda$DM even when the PBHs contribute $f_{\rm PBH}M_{\rm PBH}\ll30M_\odot$. \textcolor{black}{Blue dash-triple-dotted lines show ${\cal R}_V$ for the marked values of $v_{\rm d,rec},z$.}}
\label{fig:fig3}
\end{figure}

The advection rate $\bar{{\cal A}}_{\cal K} = - n_v \sigma_v^2/r$ is independent of $z$ in the Einstein-deSitter regime \textcolor{black}{and is added to the reduction due to expansion given by the first term on the RHS of eq. \ref{eq:v_final}. The expansion term equalizes the baryon-DM velocities at the (reduction) rate of ${\cal R}_V\equiv \frac{(1+z)}{(1+z_{\rm rec})} v_{\rm d, rec}/t_{\textcolor{black}{\rm cosm}}(z)$. The comparison between the two is shown in Fig. \ref{fig:fig3} assuming, as an example, the initial $v_{\rm d}$ at recombination to be given by $\sigma_v\simeq18$km/sec at $z=1000$. One can see that the advection rate from PBH-DM dominates for scales corresponding to those relevant for the formation of first collapsed structures already at $z\lsim 50$.
}
The precise numerology may be affected by the following:  1) $n_v$ changes in the non-linear regime at scales below $r(\delta_{\rm m,rms}\!\!=\!\!1)\simeq10^{-3} (f_{\rm PBH}M_{\rm PBH}/30M_\odot)^{1/3}[(1+z)/1000]^{-2/3}h^{-1}$Mpc, and 2) scale-dependent growth due to post-linear corrections; Figs.2,3 of \cite{Peacock:1996} show that these corrections are small for white-noise power at $\delta_{\rm m,rms}<1$. The first of these would decrease the effective advection somewhat by reducing the effective $\sigma_v$, while the second would increase it by lowering $n_v$.

Fig. \ref{fig:fig3} shows ${\cal A_K}$ from the $\Lambda$DM, PBH DM power components assuming $f_{\rm PBH}M_{\rm PBH}=3,30M_\odot$. 
As a consequence of the larger coherence of $\Lambda$DM velocity field compared to its density field, the advection rate, with ${\cal A_K}_{\rm,PBH}\propto f_{\rm PBH}M_{\rm PBH}$, is controlled predominantly by the PBH power component even at scales where the overall density field is already dominated by the $\Lambda$DM power. The red line appears in agreement with the  formalism in \cite{Tseliakhovich:2010}, which, however, is not applicable in the presence of the PBH-DM power component. The advection on small scales relevant to first source formation is driven by the PBH component even when $f_{\rm PBH}M_{\rm PBH}\ll30M_\odot$.



\section{Formation of SMBHs and galaxies in the presence of advection}

Fig. \ref{fig:fig3} shows that if PBHs make up the DM, the advection makes  baryons comoving with DM quickly after recombination with ${\cal A_K}\gsim10-100$km/sec/Gyr. Baryons and DM participate in formation/evolution of the same DM halos at the epochs they separate from the comoving frame and collapse without delay, i.e. the normal evolution for growth/collapse of density fluctuations applies \cite{Press:1974}. At the same time the same granulation component ensures an early collapse of first haloes and potentially early formation of compact objects.

Of specific relevance is the existence of SMBHs implied by QSO observations    
deep inside the reionization epoch. The following are noteworthy here in increasing $z$: 1) an ultraluminous quasar with $M_{\rm SMBH}\simeq 1.3 \times 10^{10}M_\odot$ at $z=6.3$ \citep{Wu:2015}, 2) a QSO at $z=7.1$ implying  $M_{\rm SMBH}\simeq 2 \times 10^{9}M_\odot$ \citep{Mortlock:2011}, and 3) a QSO at $z= 7.5$ with $M_{\rm SMBH}\simeq 8 \times 10^{8}M_\odot$ \citep{Banados:2018}. Their implications are significant for standard CDM model, although models have been proposed how to form them at $z\sim 12-15$ while reproducing CIB constraints (\cite{Yue:2013}, cf. \cite{Ricarte:2019}).  The difficulty stems from the limited power in the $\Lambda$DM component (Fig. \ref{fig:fig1},right) \cite{Efstathiou:1988,Kashlinsky:1993} and the advection efficiency from this component \citep[e.g.][]{Tanaka:2013}. Fig. 2 of \cite{Banados:2018} shows that, for the Eddington accretion rates, the required SMBH seed masses must be already of order a few $10^3M_\odot$ at $z\gsim 50$.



Formation of compact objects in primordial composition haloes happens if baryons there can cool and maintain certain temperatures \citep[e.g.][ and refs therein]{Bromm:2013}. This marks the critical halo masses for gas to collapse at $z$: in haloes where no H$_2$ formed the temperature would be $T\sim T_4\equiv 10^4$K, if H$_2$ formed it can reach $T\sim T_3\equiv 10^3$K. For the gas to collapse, pressure gradients must be less than gravity, defining haloes with masses $M(z)\gsim\left[\frac{4\pi(1+\delta_{\rm col})}{3}\right]^{-1/2} \left(k_{\rm B}T/m_{\rm p}G\right)^{3/2}[\bar\rho_{\rm m}(z)]^{-1/2}$, where $\delta_{\rm col}=1.68$. (This assumed $T$=const and, given its approximate values in the presence of the coolants, omitted factors of molecular weight and the slope of the pressure gradient). This delineates two critical total halo masses, $M_{4,3}$ corresponding to $T_{4,3}$ at $z$; $M_3(z)\simeq 10^6\left(\frac{1+z}{20}\right)^{-3/2}M_\odot$, $M_4(z)=(T_4/T_3)^{3/2}M_3(z)$. The halo gas mass will be a factor of $\Omega_{\rm b}/\Omega_{{\rm m},0}\sim 0.15$ smaller. H$_2$ can be destroyed by the Lyman-Werner (LW) radiation from the LIGO-type PBHs as the gas accretion onto them would lead to multi-temperature accretion disks emitting at $T_{\rm acc.disk}\sim (M_{\rm PBH}/M_\odot )^{-1/4}$keV, which may be important in any modeling involving detailed structure and compact objects formation in the case of the PBHs, with $M_4$ providing a more reliable estimate.

\begin{figure}[h!]
\includegraphics[width=4in]{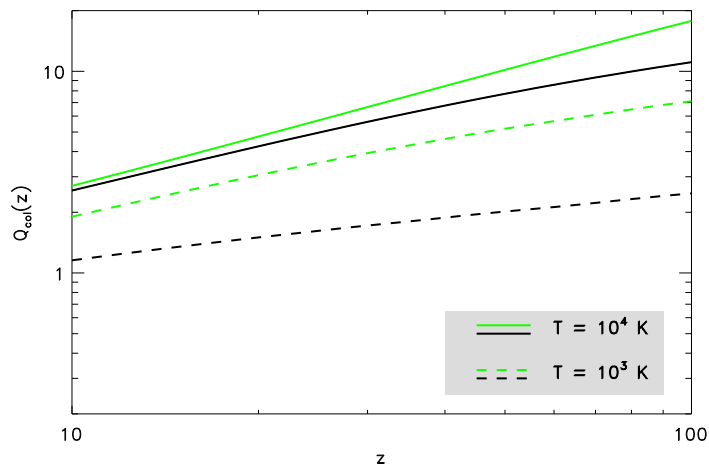}
\caption{Number of standard deviations, $Q_{\rm col}$, for halos to collapse at $z$ and cool their baryons to the given temperature. \textcolor{black}{Black(green) lines are for $f_{\rm PBH}M_{\rm PBH}=30(3)M_\odot$ and $Q_{\rm col}\propto [f_{\rm PBH}M_{\rm PBH}]^{-1/2}$ over most of the range. Granularity would also introduce non-Gaussianity \cite{Inman:2019} lowering $Q_{\rm col}$.} 
}
\label{fig:fig4}
\end{figure}

We evaluate the halo collapse likelihoods, assuming it leads to compact object's formation from the baryonic gas provided it can cool to the required $T$ and that the advection due to PBH is efficient at equalizing the DM and baryon velocities. For a Gaussian density field the probability for halo of total mass $M$ to collapse at $z$ is ${\cal P}_M(z)=\frac{1}{2}$erfc$(\frac{Q_{\rm col}}{\sqrt{2}})$ with $Q_{\rm col}\equiv \frac{\delta_{\rm col}}{\delta_{\rm m,rms}(M,z)}$ \citep{Press:1974}. At $Q_{\rm col}^2\gg1$ this reduces to ${\cal P}_M(z)\simeq \frac{1}{\sqrt{2\pi} } Q_{\rm col}^{-1}\exp(-\frac{1}{2}Q_{\rm col}^2)$ \citep{Kashlinsky:1993}; this last expression is already within $<15\%$ of the true ${\cal P}_M$ for $Q_{\rm col}\geq 2.25$. Fig. \ref{fig:fig4} shows $Q_{\rm col}(z)$ for the two cooling regimes. In the PBH-DM paradigm one can have a reasonable abundance of haloes, say of $Q_{\rm col}(z)\lsim 3-6$, with the gas collapsing on mass-scales up to $\lsim 10^5M_\odot$ at $z\lsim 30-50$, if only a subset of these systems subsequently forms seed BHs of $M_{\rm BH, seed}\sim 10^{3-4}M_\odot$ via e.g. any of the numerous mechanisms suggested \citep[e.g.][and refs therein]{Latif:2016}.  \textcolor{black}{A number of models, applicable here, have been developed for formation of SMBHs inside collapsed halos at high $z$. The proposed mechanisms are a result of stellar dynamical relaxation processes, typically involving evolution of dense stellar systems, coupled with gas collapse and dynamic instabilities discussed by e.g. \cite[][]{Begelman:1978,Kashlinsky:1983,Begelman:2006}. Since $Q_{\rm col}\propto [f_{\rm PBH}M_{\rm PBH}]^{-1/2}$, to form collapsed halos with $T_3$ by $z\sim 40$ would require $f_{\rm PBH}M_{\rm PBH}\gsim 3M_\odot$ setting $Q_{\rm col}\lsim6$.} Because of the advection efficiency {\it and} the extra power of the PBH-DM component the SMBHs appear to support the PBH-DM connection conjecture.

In the PBH-DM paradigm the advection rate as shown in Fig. \ref{fig:fig3} for $f_{\rm PBH}M_{\rm PBH}=30\textcolor{black}{,3}M_\odot$ reaches minimum around dark matter mass scales $M_{\rm d}\sim10^9\textcolor{black}{,5\times10^7}M_\odot$ and then rises again. This may require modifications and additional considerations in various discussions involving subsequent structure formation \citep[e.g.][]{Carr:2018}. Naively this implies that there may be a pause after the first collapse era and resurgence of collapse and luminous source formation around masses of order modern galaxies. 
After the pause the gas would collapse to form the presently observed galaxy morphology depending on the halo spin \citep{Kashlinsky:1982}.  \textcolor{black}{This can explain the existence of early-type galaxies with already established morphology and $\gsim 3.5$Gyr-old stellar populations at $z\sim 1.5-2$ \cite{Dunlop:1996,Saracco:2009,Renzini:2006}.}

%
\acknowledgements Fernando Atrio-Barandela and Alexandre Vassilkov are thanked for discussions on cosmological aspects and advection hydrodynamics respectively. Support from NASA/12-EUCLID11-0003 ``LIBRAE: Looking at Infrared Background Radiation Anisotropies with Euclid" project 
is acknowledged. 



%

\end{document}